\documentclass{PoS}

\usepackage{amsmath}
\usepackage{MnSymbol}

\newcommand\tabvsptop{\rule{0pt}{2.6ex}}

\newcommand{\refeq}[1]{Eq.~(\ref{eq:#1})}

\newcommand{\reffig}[1]{Fig.~\ref{fig:#1}}

\newcommand{\reftab}[1]{Table~\ref{tab:#1}}

\def \Op{{\cal O}}

\title{Theory status of $b\to s\,\ell^+\ell^-$ decays and their combined analysis}

\ShortTitle{$b\to s\,\ell^+\ell^-$: Theory \& combined analysis}

\author{\speaker{Christoph Bobeth}\\
        Excellence Cluster Universe\\
  Technische Universit\"at M\"unchen\\
  D-85748 Garching, Germany \\
        E-mail: \email{christoph.bobeth@ph.tum.de}}

\abstract{
The experimental information on $b\to s\,\ell^+\ell^-$ decays has grown enormously
during the last couple of years due to BaBar, Belle, CDF and LHCb. Especially,
exclusive modes $B\to K^{(*)}\ell^+\ell^-$, which have the largest rate and are 
easily accessible experimentally, provide a variety of observables which 
constrain non-standard interactions that would affect these flavour-changing neutral
current decays beyond the Standard Model. Nowadays, theoretical predictions focus
on low- and high dilepton invariant mass regions, where expansions in 
$\Lambda_{\rm QCD}/m_b$ and form factor symmetries provide means to identify
optimised observables. The first experimental results of non-optimised observables
have stimulated first global analysis of $b\to s\,\ell^+\ell^-$ decays in 
combination with $b\to s\gamma$ and $B_s\to \mu^+\mu^-$. Such global analysis
are now ready to be applied to include high-statistics results from LHCb and 
Super-Flavour factories which are about to come within the next years.
}

\FullConference{The XIth International Conference on Heavy Quarks and Leptons,\\
		June 11-15, 2012\\
		Prague, Czech Republic}

\begin{document}

In the past the experimental program of quark flavour physics adressed primarily 
the exploration of CP-violation in the $B$-system and the tightly related picture
of quark flavour mixing in the Standard Model (SM) represented by the unitary 
Cabibbo-Kobayashi-Maskawa (CKM) matrix. However, the ever increasing luminosity
allows nowadays also to explore flavour-changing neutral current (FCNC) decays
of $b$-hadrons which are loop-suppressed in the SM and therefore have an enhanced
sensitivity to non-standard virtual contributions. They test the SM at the 
loop-level and constitute indirect searches for non-standard effects. Consequently,
precise experimental measurements are needed and a good control is required over
theoretical uncertainties.

The class of FCNC decays, mediated by $b\to s\,\ell^+\ell^-$ $(\ell = e,\,\mu,\,\tau)$
is a phenomenologically rich sub-class with larger branching fractions of $\sim 
{\cal O}(10^{-6})$ (in the SM), compared to the CKM-suppressed $b\to d\,\ell^+\ell^-$
decays $|V_{td}/V_{ts}|^2 \sim {\cal O}(10^{-2})$. It comprises inclusive and exclusive
semi-leptonic decays $B_{u,d} \to (X_s,\, K,\, K^*)\, \ell^+\ell^-$, $B_s \to (f_0,\, \phi)
\, \ell^+\ell^-$, $\Lambda_b \to \Lambda \,\ell^+\ell^-$ as well as the purely 
leptonic $B_s\to \ell^+\ell^-$ decay. Further channels with excited $K^*$ have
been discussed in the literature. Within the last few years four experimental
collaborations analysed some of these channels with number of events in the range of $150$ to
$250$ for BaBar \cite{:2012vw}, Belle \cite{:2009zv} and CDF \cite{Aaltonen:2011qs} 
and about $1000$ events at LHCb \cite{Aaij:2011aa} as listed in \reftab{numofevents}.
The results based on the final data set have been released this year by BaBar and
announced this summer by CDF, whereas Belle's results are based on a partial data set.
By now LHCb dominates statistically, adding about 2 fb$^{-1}$ of data this year 2012
and possibly another 4 fb$^{-1}$ until the year 2018 before a long shut-down. The 
Super-Flavour factories Belle II \cite{Aushev:2010bq} and SuperB \cite{O'Leary:2010af}
will be able to collect data sets with about 10000 -- 15000 events \cite{Bevan:2011br}. 

Theoretical predictions of $b\to s\, \ell^+\ell^-$ decays are obtained using
the effective theory of $\Delta B = 1$ decays of the electroweak interaction
of the SM. It provides the universal starting point for the calculation of
observables of inclusive and exclusive decays. The short-distance information
at the electroweak sale $\mu \sim M_W$ of the order of the mass of the $W$-boson
are contained in effective coupling constants $C_i$ (Wilson coefficients) whereas
flavour-changing interactions of $b\to s$ are described by one dimension five $b\to s\gamma$
operator $\Op_7$ and two dimension six $b\to s\, \ell^+\ell^-$ operators $\Op_{9,10}$.
Due to operator mixing, additional 4-quark operators have to be included, which
are the current-current operators $\Op_{1,2}^{u,c}$, the QCD-penguin operators 
$\Op_{3,4,5,6}$ and the chromo-magnetic dipole operator $\Op_8$. The effective 
Hamiltonian reads \cite{Chetyrkin:1996vx, Bobeth:1999mk}
\begin{align}
  \label{eq:Heff}
  {\cal{H}}_{\rm eff} & = -\frac{4 G_F}{\sqrt{2}} V_{tb}^{} V_{ts}^\ast
     \left( {\cal{H}}_{\rm eff}^{(t)} + \hat{\lambda}_u {\cal{H}}_{\rm eff}^{(u)} \right) + \mbox{h.c.}, &
    \hat{\lambda}_u & = V_{ub}^{} V_{us}^\ast/V_{tb}^{} V_{ts}^\ast ,
\end{align}
\begin{align}
  \label{eq:Heff:parts}
  {\cal{H}}_{\rm eff}^{(t)} & =
    C_1^{} \Op_1^c + C_2^{} \Op_2^c + \sum_{3 \leq i} C_i^{} \Op_i^{}, &
  {\cal{H}}_{\rm eff}^{(u)} & =
    C_1^{} (\Op_1^c - \Op_1^u) + C_2^{} (\Op_2^c - \Op_2^u)
\end{align}
with 
\begin{align}
  \Op_{7} & = \frac{e}{(4 \pi)^2} m_b \left(\bar{s}\, \sigma^{\mu\nu} P_R\, b \right) F_{\mu\nu}, &
  \Op_{9\, (10)} & = \frac{e^2}{(4 \pi)^2} \left(\bar{s}\, \gamma^\mu P_L\, b\right)
                  \left(\bar\ell\, \gamma_\mu (\gamma_5)\, \ell\right)
\end{align}
where the Wilson coefficients are renormalised in the $\overline{\mbox{MS}}$-scheme
and evaluated at the renormalisation scale $\mu \sim m_b$ of the order of the $b$-quark
mass. They have been calculated in the SM up to the next-to-next-to-leading order (NNLO)
in QCD. At higher order in the electromagnetic coupling also QED-penguin operators have
been considered for the inclusive decay \cite{Bobeth:2003at}. Within extensions
of the SM, new contributions arise $C_i^{} \to C_i^{\rm SM} + C_i^{\rm NP}$ and additional
$(\bar{s}\ldots b)(\bar\ell\ldots \ell)$ operators can contribute which have zero or negligible
Wilson coefficients in the SM. For example right-handed currents give rise to 
chirality-flipped $\Op_{7',9',10'}$ obtained by the interchange $P_L \leftrightarrow P_R$.
Scalar and pseudo-scalar operators $\Op_{S,S',P,P'}$ can have enhanced contributions
from neutral Higgs-penguin or box-type diagrams, where the latter give also rise to tensor
operators $\Op_{T,T5}$. There are also non-standard scenarios which give rise to FCNC's
at tree-level, such as LeptoQuark's or extensions with non-unitary quark mixing matrices.
CP violation is suppressed in the SM in $b\to s$ transitions due to the smallness of
$\mbox{Im}[\hat{\lambda}_u]$ which is doubly Cabibbo-suppressed.

\begin{table}
\centering
\begin{tabular}{l|c|c|c|c}
 \# of evts & BaBar 2012 & Belle 2009 & CDF 2011 & LHCb 2011
\\[0.0cm]
  & 471 M $\bar{B}B$ 
  & $605$ fb$^{-1}$ 
  & $6.8$ fb$^{-1}$ 
  & $1$ fb$^{-1}$
\\
\hline
\tabvsptop
  $B^0\to K^{*0}\, \ell^+\ell^-$ 
  & $137 \pm 44^\dagger$ 
  & $247\pm54^\dagger$ 
  & $164 \pm 15$ 
  & $673 \pm 30$
\\[0.1cm]
  $B^+\to K^{*+}\, \ell^+\ell^-$ 
  &  
  & 
  & $20 \pm 6$
  & $76 \pm 16$
\\[0.1cm]
  $B^+\to K^+ \, \ell^+\ell^-$ 
  & $153 \pm 41^\dagger$ 
  & $162 \pm 38^\dagger$ 
  & $234 \pm 19$ 
  & $1250 \pm 42$
\\[0.1cm]
  $B^0\to K^0_S\, \ell^+\ell^-$
  & 
  & 
  & $28 \pm 9$ 
  & $60 \pm 19$
\\[0.1cm]
\hline
\tabvsptop
  $B_s \to \phi\, \ell^+\ell^-$
  & 
  &
  & $49 \pm 7$ & $77 \pm 10$
\\[0.1cm]
\hline
\tabvsptop
  $\Lambda_b \to \Lambda \,\ell^+\ell^-$
  & 
  & 
  & $24 \pm 5$ 
  &
\end{tabular}
\caption{ \label{tab:numofevents}
  The number of events for various exclusive $b\to s\,\ell^+\ell^-$ decays (CP- and
  lepton-flavour $(\ell = e,\, \mu)$ averaged) obtained by BaBar \cite{:2012vw},
  Belle \cite{:2009zv}, CDF \cite{Aaltonen:2011qs} and LHCb \cite{Aaij:2011aa}. 
  The $q^2$ region around $J/\psi$ and $\psi'$ have been vetoed. 
  ${}^\dagger$ An unknown mixture of $B^0$ and $B^\pm$.
}
\end{table}

The Wilson coefficients of the loop-induced $b\to s\gamma$ and $b\to s\,\ell^+\ell^-$ 
SM operators $\Op_{7,9,10}$ -- and potentially non-standard operators $\Op_i$ with 
$i = 7',9',10',S,S',P,P',T,T5$ -- are of great interest for indirect searches of 
signatures beyond the SM. They constitute the numerically leading contribution for
large parts in the kinematic region of the dilepton invariant mass $q^2$ in most of
the observables. However, when $q^2$ approaches production thresholds of
$q\bar{q}$-resonances, 4-quark operators $b\to s\, q\bar{q}$ induce an additional
interfering amplitude $b\to s\, (q\bar{q}) \to s\, \ell^+\ell^-$ which involves
nonperturbative dynamics that are theoretically not well under control. Especially
the current-current operators $\Op_{1,2}^q$ with $q=c$ result in large peaking
backgrounds $b\to s\, J/\psi$ and $b\to s\, \psi'$ with branching fractions of
$\Op(10^2)$ larger than the once from $\Op_{7,9,10}$ and are vetoed in the 
experimental analysis. Analogous contributions for $q = u$ are suppressed by
$\hat{\lambda}_u$ whereas QCD-penguin operators have tiny Wilson coefficients.

Exclusive decays are currently available with high-statistics for the two most
prominent decays $B^+\to K^+\ell^+\ell^-$ and $B^0 \to K^{*0} (\to K^+\pi^-)\,
\ell^+\ell^-$. In comparison, the current experimental precision of inclusive
decays can not compete and future measurements at Super-Flavour factories have
to be awaited. In view of this, the theoretical status will be discussed only
for exclusive decays in the following.

\section{Exclusive decays}

Two distinct theoretical methods have been applied in order to calculate observables
of exclusive decays $B\to M\, \ell^+\ell^-$, where $M$ denotes light mesons 
$K,\, K^*$, for the two regions of dilepton invariant mass below and above the
two narrow $c\bar{c}$-resonances $J/\psi$ and $\psi'$. These methods are based
on the different kinematical limits of large hadronic recoil (of $M$) at low-$q^2$
and low hadronic recoil at high-$q^2$ which allow for expansions in 
$\lambda \equiv \Lambda_{\rm QCD}/m_b$.
 
At low-$q^2$, contributions due to $b\to s\,q\bar{q}$ ($q = u,d,s,c$) 4-quark
operators and the $b\to s \, gluon$ operator $\Op_8$ are treated within QCD 
factorization (QCDF) \cite{Beneke:2001at} using the large energy limit of the 
recoiling meson $M$. It allows to include NLO corrections in the strong coupling 
$\alpha_s$ and effects of spectator-quark
scattering. The amplitudes factorise schematically into perturbatively calculable
quantities $C$ and $T$ and nonperturbative objects, form factors $\xi$ and
meson-distribution amplitudes $\phi_{B,M}$,
\begin{align}
  \label{eq:lowQ2:amp}
  A &Ê\sim 
    C\, \xi + \phi_B \otimes T \otimes \phi_M 
  + \Op\left(\lambda\right)
\end{align}
where $\otimes$ implies a convolution over the according momentum fractions.
Notably, the large energy symmetry relations for heavy-to-light form factors
allow to reduce seven $B\to V$ form factors to two universal $\xi_{\perp,\parallel}$
and the three $B\to P$ form factors to one $\xi_P$ \cite{Charles:1998dr} which
are valid up to order $\lambda$ corrections and constitute the first part
of lacking terms in \refeq{lowQ2:amp}. The corresponding $\alpha_s$ corrections
at leading order in $\lambda$ have been included in $C$ and $T$ 
\cite{Beneke:2000wa}. For example the three $K^*$-transversity amplitudes in 
$B\to K^*\ell^+\ell^-$ decays have a simple form at leading order in
$\lambda$ and $\alpha_s$~\cite{Kruger:2005ep}
\begin{align}
  \label{eq:lowQ2amp}
  A_{\perp,\parallel}^{L,R} & 
  \sim \pm \,C_\perp^{L,R} \times \xi_\perp + \Op\left(\alpha_s, \lambda\right), &
  A_{0}^{L,R} & 
  \sim C_\parallel^{L,R} \times \xi_\parallel + \Op\left(\alpha_s, \lambda\right), 
\end{align}
where the two short-distance coefficients are a linear combination of $C_{7,9,10}$
\begin{align}
  C_\perp^{L,R} & 
   = (C_9 \mp C_{10}) + \frac{2 m_b M_B}{q^2} C_7, &
  C_\parallel^{L,R} & 
   = (C_9 \mp C_{10}) + \frac{2 m_b }{M_B} C_7. &
\end{align}
Besides the form factor symmetries, a second source of lacking sub-leading
corrections in $\lambda$ are due to the expansions of the amplitude 
itself. They involve also divergent contributions of distribution amplitudes 
$\phi_M$ at the sub-leading order which introduce a model-dependence. This affects
especially the isospin-asymmetry which arises due to differences in spectator
interactions \cite{Feldmann:2002iw}. Additionally, soft-gluon effects from
$c\bar{c}$-resonances due to current-current operators $\Op_{1,2}^c$ have 
been calculated within a non-local OPE \cite{Khodjamirian:2010vf}
for the tails at $q^2$ below the resonances. They can change the rate
up to $(10 - 20) \%$ for $q^2$ values of interest $\sim 6$~GeV$^2$, rising 
further for values even closer to the resonances.
 
At high-$q^2$, a local operator product expansion (OPE) can be applied to the
contributions of 4-quark operators due to the hard momentum $\Lambda_{\rm QCD}
\ll q^2 \sim m_b^2$ \cite{Grinstein:2004vb, Beylich:2011aq} which is passing
through the $q\bar{q}$-resonance. Now the $K^*$-transversity amplitudes
depend only on one coefficient $C^{L,R}$ \cite{Bobeth:2010wg}
\begin{align}
  \label{eq:highQ2amp}
  A_i^{L,R} & \sim
    C^{L,R} \times f_i 
    + C_7 \times \Op\left(\lambda\right) 
    + \Op\left(\lambda^2\right), & 
  C^{L,R} & 
   = (C_9 \mp C_{10}) + \kappa \frac{2\, m_b^2}{q^2} C_7.
\end{align}
The well-known Isgur-Wise form factor relations \cite{Isgur:1990kf}, improved by
the inclusion of QCD corrections $\kappa$ \cite{Grinstein:2004vb}, can be used to
eliminate the three tensor form factors by the vector and axial-vector form factors.
The according linear combinations are denoted by $f_i$ $(i = 0, \perp, \parallel)$ 
\cite{Bobeth:2010wg}. Due to the local OPE $B\to M$ form factors arise as the only
nonperturbative objects, which are at the lowest order (dimension 3) the usual QCD
form factors \cite{Grinstein:2004vb, Beylich:2011aq}. The use of the form factor
relations introduces an uncertainties of order $\lambda$ which is $\propto C_7$.
However, since the numerically leading term is dominated by $|C_{9,10}| \sim 4.2$ 
in comparison to $|C_7| \sim 0.3$ (in the SM), an additional numerical suppression
of $|C_7/C_{9,10}|\sim 0.1$ arises. In the OPE dimension four terms are absent
such that sub-leading contributions to the amplitude are suppressed by $\lambda^2$.
At higher orders in the OPE new form factors of the higher dimensional operators
enter. All such form factors can be calculated at high-$q^2$ in principle on the
lattice due to the low recoil of $M$. The NLO $\alpha_s$ corrections to the
dimension three term are known as well \cite{Seidel:2004jh} and lead to small
renormalisation scale dependences. Finally, duality violating contributions to the
OPE have been estimated based on a model and found to be of a few percent at the
level of the rate when integrating over sufficiently large $q^2$-bins 
\cite{Beylich:2011aq}.

The main uncertainties in predictions of exclusive decays are due to form factors
and lacking sub-leading contributions in the power expansions in $\lambda$.
At high-$q^2$ sub-leading contributions are calculable and their omission less
problematic due to the stronger suppression \refeq{highQ2amp} compared to \refeq{lowQ2amp}.
Moreover, the according form factors of higher-dimensional operators can be 
calculated in principle on the lattice. Contrary, at low-$q^2$, no approach is 
known to the arising divergences in convolutions of distribution 
amplitudes at sub-leading order in QCDF, which results in a larger theoretical
uncertainty in this region. Currently, the form factors are known mainly from
light-cone sum rule (LCSR) calculations which are restricted to low-$q^2$
\cite{Ball:2004ye, Khodjamirian:2010vf}. At high-$q^2$ form factors have been
calculated for $B\to K$ on the lattice in quenched approximation \cite{Becirevic:2006nm,
Becirevic:2012fy}, whereas only preliminary unquenched results for $B\to K^{(*)}$
are reported without final error estimates \cite{Liu:2011ra}. Therefore, currently
all predictions at high-$q^2$ rely on extrapolations of the LCSR form factor
results from the low-$q^2$. Unquenched lattice results should become available
in the close future for $B\to K$ and $B\to K^*$.

\section{Optimised Observables}

Both decays, $B^+\to K^+ \ell^+\ell^-$ and $B^0 \to K^{*0}(\to K^+\pi^-)\ell^+\ell^-$,
allow to measure several observables in their angular distributions. For the 3-body 
final state in $B^+\to K^+ \ell^+\ell^-$ this is the angle $\theta_\ell$ between the
$\ell^-$ momentum and the direction of flight of the meson $M$ in the $(\ell^+\ell^-)$ 
center of mass (CMS) frame. In the 4-body final state $B^0 \to K^{*0}(\to K^+\pi^-)\ell^+\ell^-$
two additional angles exist. These are an analogous angle $\theta_K$ between the 
Kaon momentum and the $K^*$ momentum in the $(K\,\pi)$ CMS frame and the angle $\phi$
spanned by the two decay planes of the $(\ell^+\ell^-)$ and $(K\, \pi)$ systems. The
intermediate $K^*$ is assumed to be on-shell, such that the $(K\, \pi)$ invariant
mass is fixed to the mass of the $K^*$ and the narrow width approximation is used
frequently. A very recent work addressed the latter issue \cite{Becirevic:2012dp}
by including a finite width for the $K^*$ and two additional scalar resonances 
finding a non-negligible impact at low-$q^2$ depending on the observable and the
value of $q^2$.

Currently, for $B^+\to K^+ \ell^+\ell^-$ the branching ratio $Br$, the lepton 
forward-backward asymmetry $A_{\rm FB}$ and the isospin asymmetry $A_I$ have been
measured in $q^2$-bins covered by the theoretical methods described above, whereas 
for $B^0 \to K^{*0} \ell^+\ell^-$ these are $Br$, $A_{\rm FB}$, $A_I$,
the longitudinal $K^*$ polarization fraction $F_L$, and further observables in the 
$\phi$-distribution by LHCb and CDF: $S_{3,9}$ and $A_T^{(2)}$, $A_{im}$. 

In $B^+\to K^+ \ell^+\ell^-$ the angular distribution w.r.t. $\cos\theta_\ell$ allows to
measure the lepton $A_{\rm FB}$ and the observable $F_H$ \cite{Bobeth:2007dw, Bobeth:2011nj}.
The first is very sensitive to scalar and tensor $(\bar{s}\ldots b)(\bar\ell\ldots \ell)$ 
operators which are absent in the SM and similarly for $F_H$ \cite{Bobeth:2007dw}. 
In the presence of chirality-flipped operators $\Op_{7',9',10', S', P'}$ the combination
$(C_i + C_{i'})$ enters all observables, contrary to $B_s \to \ell^+\ell^-$ which
depends on $(C_i - C_{i'})$ (for $i = 10, S, P$) and $B\to K^*\ell^+\ell^-$, which
depends on both combinations.

The structure of the $K^*$-transversity amplitudes (\ref{eq:lowQ2amp}) and 
(\ref{eq:highQ2amp}) has phenomenological interesting con\-se\-quences for the
4-body final state in $B \to K^*(\to K\pi)\ell^+\ell^-$ decays. Its angular analysis
offers a large number of angular observables $J_i(q^2)$ ($i = 1s,\, \ldots\, 9$) 
\cite{Kruger:1999xa}, such that suitable combinations of $J_i(q^2)$ could
be identified which exhibit a reduced hadronic uncertainty and enhanced sensitivity
to short-distance couplings of the SM and scenarios beyond. These ``optimized
observables'' are at low-$q^2$ $A_T^{(2,3,4,5,{\rm re, im})}$ and $P_{4,5,6}$ 
\cite{Kruger:2005ep, Egede:2008uy, Matias:2012xw} whereas at high-$q^2$ $H_T^{(2,3,4,5)}$
\cite{Bobeth:2010wg}. Further observables have been identified in the presence of
scalar operators at low-$q^2$ \cite{Matias:2012xw}. Additionally, at high-$q^2$ 
also combinations are known which do not depend on the short-distance couplings 
(mostly in the SM operator basis) \cite{Bobeth:2010wg} and allow to probe the
form factor shapes with data \cite{Hambrock:2012dg, Beaujean:2012uj}. CP asymmetric
combinations with reduced hadronic uncertainties have been also found at low-$q^2$
\cite{Kruger:2005ep, Egede:2008uy} and high-$q^2$ \cite{Bobeth:2011gi}. The
sensitivity to $B_s$-mixing parameters $\phi_s$ and $\Delta \Gamma_s$ in
time-integrated CP asymmetries of $B_s \to \phi(\to K^+K^-)\,\ell^+\ell^-$ turns 
out to be small \cite{Bobeth:2011gi, Bobeth:2008ij}. The $J_i(q^2)$ normalised to
the decay rate and the associated CP-asymmetries have been also studied
model-independently and model-dependently in great detail \cite{Bobeth:2008ij}.
At the moment no experimental measurements are available for the optimised
observables, except for $A_T^{(2)}$ from CDF \cite{Aaltonen:2011qs}.

\section{Global analysis}

Currently, global analysis of radiative, semi-leptonic and leptonic decays combine
available data for inclusive $B\to X_s (\gamma,\, \ell^+\ell^-)$, exclusive
$B\to K \ell^+\ell^-$, $B\to K^*(\gamma,\, \ell^+\ell^-)$ and the leptonic 
$B_sÊ\to \mu^+\mu^-$ modes. There are model-independent studies which determine
the constraints on the Wilson coefficients for varying sets of operators or 
model-dependent studies which derive bounds on the parameters of extensions 
of the SM.

The determination of confidence or probability regions varies in all analysis.
Most simple approaches determine allowed regions of parameter space by combining
$n-\sigma$ $(n = 1, 2, 3)$ experimental and theoretical errors \cite{DescotesGenon:2011yn,
Becirevic:2012fy}. Others calculate $\chi^2$ values by combining experimental and
theoretical errors following the $R$-fit scheme \cite{Bobeth:2011gi, Bobeth:2011nj}
or different definitions \cite{Altmannshofer:2011gn, Hurth:2012jn, DescotesGenon:2012zf}. 
A third approach includes parameters associated with theoretical uncertainties as 
nuisance parameters in the fit \cite{Beaujean:2012uj}. Once more precise data will
be available, the more sophisticated methods should be used and further experimental
correlations among observables have to be included as well, a task requiring 
a close collaboration between experimental and theory sides. First dedicated 
software tools have been developed for exclusive $b\to s\, \ell^+\ell^-$ decays
\cite{EOS:website} or extended \cite{Mahmoudi:2007vz}.

\begin{figure}
\includegraphics[width=0.5\textwidth]{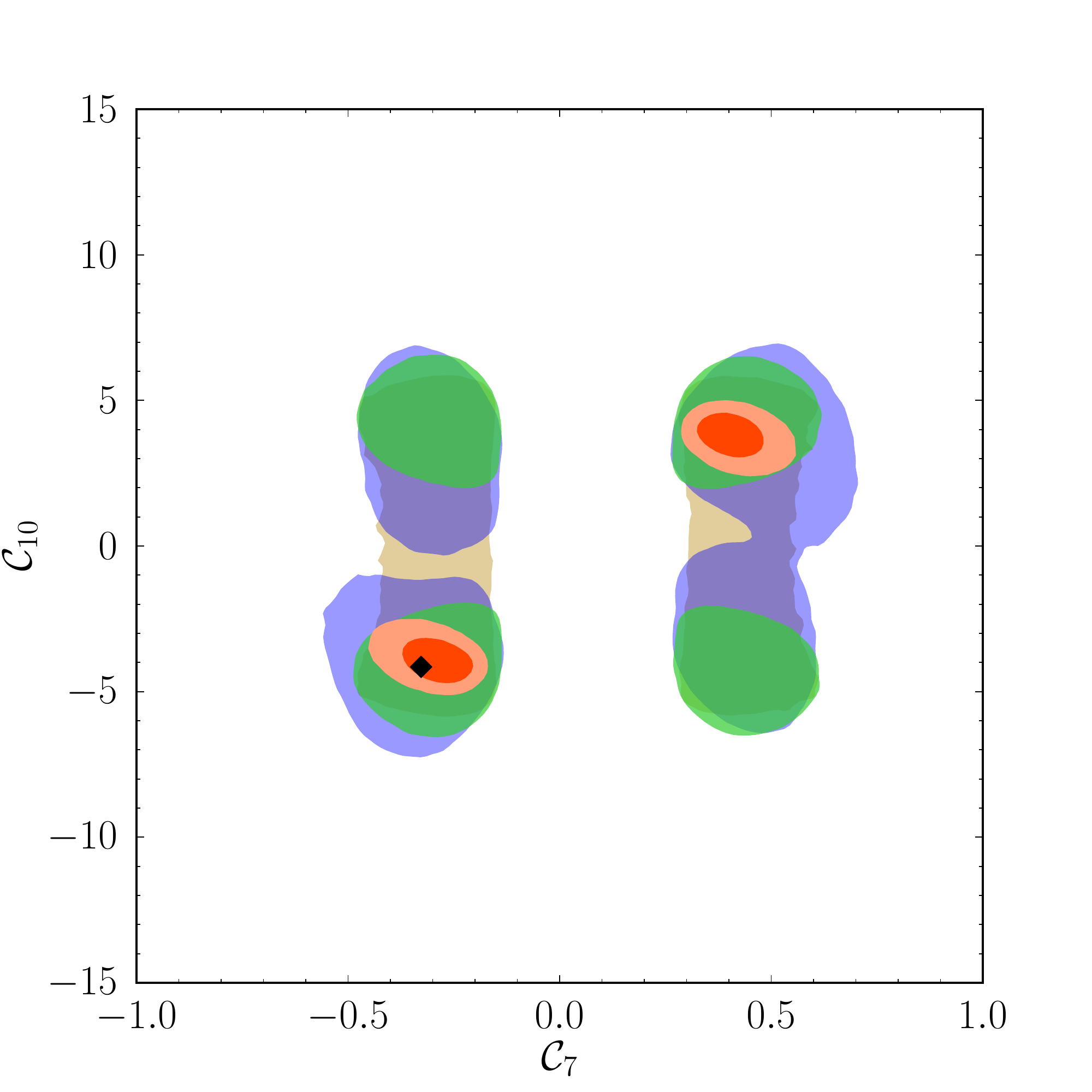}
\includegraphics[width=0.5\textwidth]{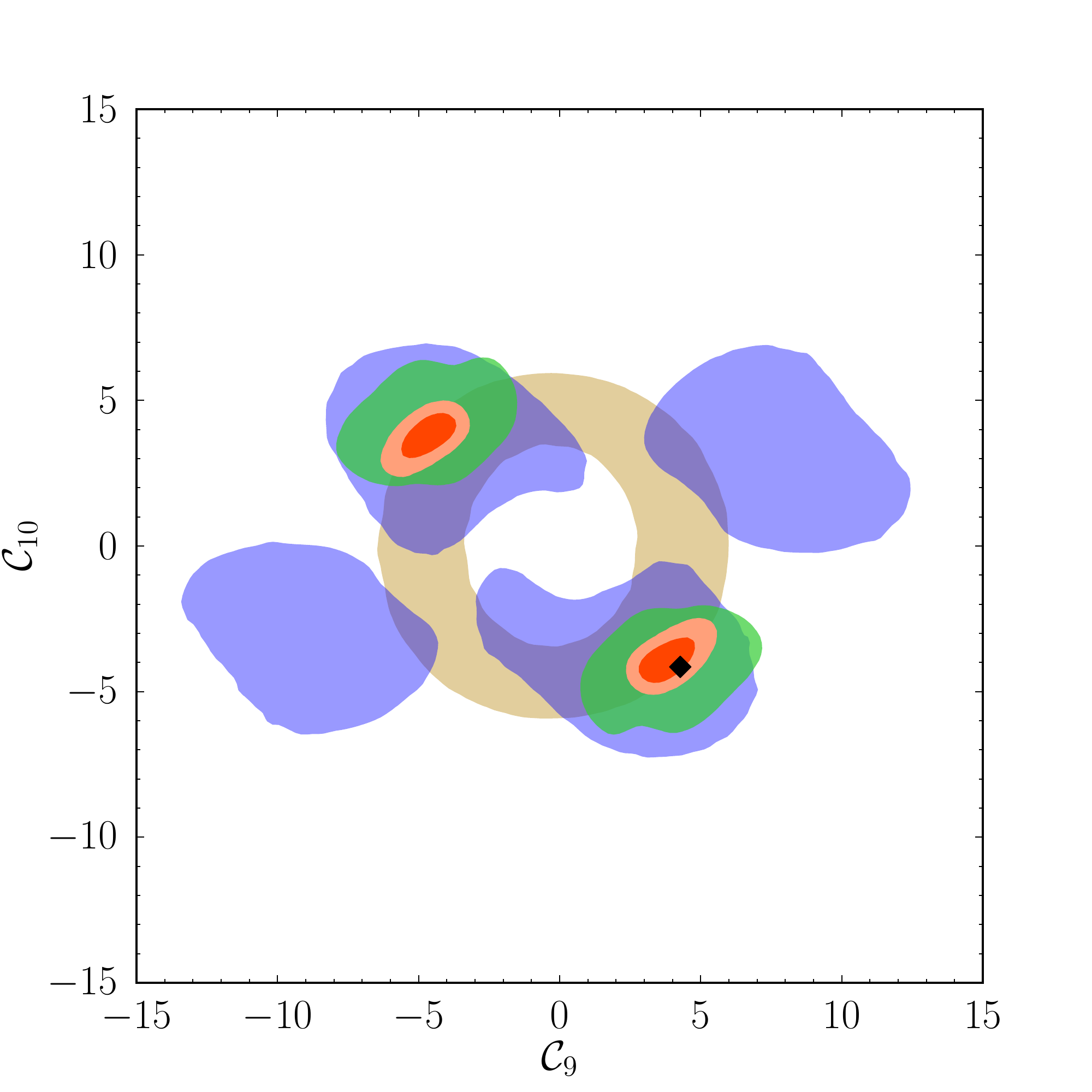}
\caption{
  The marginalized 2-dimensional 95\% credibility regions of the
  Wilson coefficients $C_{7,9,10}$ for $\mu = 4.2$ GeV are shown
  when applying the $B\to K^* \gamma$ constraints in combination with
  {\it i)} only low- and high-$q^2$ data from $B \to K\, \ell^+\ell^-$ [brown];
  {\it ii)} only low-$q^2$ data from $B \to K^* \ell^+\ell^-$ [blue];
  {\it iii)} only high-$q^2$ data from $B \to K^*\ell^+\ell^-$ [green];
  and {\it iv)} all the data, including also $B_s \to \mu^+\mu^-$
  [light red], showing as well the 68\% credibility interval [red].
  The SM values $C^{\rm SM}_{7,9,10}$ are indicated by $\filleddiamond$
  \cite{Beaujean:2012uj}.
}
\label{fig:c7910constraints}
\end{figure}

In the context of model-independent analysis, the simplest scenario assumes
new physics in $C_{7,9,10}$ which is real, i.e. involves the same CP-violation
as in the SM. The according results can be seen in \reffig{c7910constraints}
from exclusive decays only when combining $B\to K^*\gamma$, $B\to K \ell^+\ell^-$
and $B\to K^* \ell^+\ell^-$ as well as from single processes only. Two solutions
remain, which are related by a simultaneous sign-flip of all three Wilson 
coefficients compared to the SM signs. The solution with only $C_7$ flipped is
now excluded \cite{Beaujean:2012uj, Altmannshofer:2011gn}, mainly due to the
measurements of $A_{\rm FB}$. It can be also seen, that the high-$q^2$ region
plays currently a crucial role. Overall, the SM is in good agreement with the
data, but large deviations are still not excluded.

More general scenarios have been analysed too, including chirality-flipped
Wilson coefficients \cite{DescotesGenon:2011yn, Altmannshofer:2011gn, Becirevic:2012dx,
DescotesGenon:2012zf} assuming them to be real or complex. Especially the
measurement of the optimised observables will be important in order to
efficiently constrain these scenarios which currently still allow for
large deviations from SM predictions, especially in optimized observables
and CP asymmetries.

In the framework of the minimal supersymmetric SM (MSSM) the $b\to s\, \ell^+\ell^-$
transitions provide constraints on flavour-changing left-right mixing in the
up-squark-sector $(\delta^u_{23})_{LR}$, which in turn place constraints on
top-quark FCNC decays $t\to c\gamma$, $t\to cg$ and $t\to cZ$ \cite{Behring:2012mv}.
The interplay of $B_s\to \mu^+\mu^-$ at large $\tan\beta$ and angular observables
in $B\to K^* \ell^+\ell^-$ at moderate $\tan\beta$ has been investigated in
constrained scenarios such as the CMSSM and NUHM \cite{Mahmoudi:2012un}.
LeptoQuark interactions, which induce scalar and pseudo-scalar operators 
$\Op_{S,S',P,P'}$, have been constraint with recent data from $B\to (X_s, K)\, \ell^+\ell^-$,
and $B_s\to \mu^+\mu^-$ \cite{Kosnik:2012dj}.

The transition $b\to s\,\tau^+\tau^-$ is experimentally not constrained except for
the recent upper bound $Br(B^+ \to K^+\tau^+\tau^-) < 3.3\cdot 10^{-3}$ from BaBar
\cite{Flood:2010zz}. Indirect constraints might be derived using recent measurements
of life-time ratio $\tau_{B_d}/\tau_{B_s}$ which however involve some assumptions.
The resulting constraints on the Wilson coefficients of $(\bar{s}\,b)(\bar\tau\,\tau)$
operators have been discussed in \cite{Bobeth:2011st}. They do not allow for larger
modifications of the decay width difference $\Delta \Gamma_s$ of the $B_s$-meson 
in the SM than $35$\% assuming single operator dominance. The interplay of 
several operators allows for larger effects \cite{Dighe:2012df}.

With the higher statistics in experimental results and form factor predictions
from lattice the currently proposed observables and strategies of their combination
will allow to derive stronger constraints on new physics scenarios beyond
the Standard Model which will be complementary to other particle physics
sectors. Currently the SM describes the data, but in the future it can be
tested more stringently, especially with the first measurements of optimized 
observables.

\vskip 0.5cm

\noindent
{\bf Acknowledgments} $\quad$ I am indebted to the organisers of the {\em HQL-2012}
for the opportunity to present a talk and the kind hospitality in Prague. 
I thank Frederik Beaujean, Ulrich Haisch, Gudrun Hiller, Danny van Dyk and Christian 
Wacker for our fruitful collaborations. 


\end{document}